\documentclass[reprint,superscriptaddress,preprintnumbers,nofootinbib, amsmath,amssymb, prdfloatfix]{revtex4-1}
\usepackage{CJK}
\usepackage{graphicx}
\usepackage{dcolumn}
\usepackage{upgreek}
\usepackage{bm}
\usepackage[colorlinks=true,pdfstartview=FitV,breaklinks=true]{hyperref}
\usepackage[dvipsnames,table]{xcolor}
\hypersetup{urlcolor=BlueViolet,
	    citecolor=Plum,
	    linkcolor=PineGreen}
\usepackage{lipsum}	    
\usepackage{titlesec}
\titlespacing*{\section}{0pt}{0.8\baselineskip}{0.8\baselineskip}
\usepackage{times}

\begin{document}

\preprint{CPPC-2021-11}

\title{Apparent horizons of the Thakurta spacetime and the description of cosmological black holes}

\author{Archil Kobakhidze}
 \email{archil.kobakhidze@sydney.edu.au}
 \address{ 
Sydney Consortium for Particle Physics and Cosmology, \\
 School of Physics, The University of Sydney, NSW 2006, Australia 
}

 \author{Zachary S. C. Picker}
 \email{zachary.picker@sydney.edu.au}
\address{
School of Physics, The University of Sydney and \\
 ARC Centre of Excellence for Dark Matter Particle Physics, 
 NSW 2006, Australia 
}

\begin{abstract}
We discuss the validity of the Thakurta metric to describe cosmological black holes by analysing the nature of its horizon. By adopting the preferred foliation of the Thakurta spacetime associated with the Kodama time, we demonstrate that the Thakurta horizon is indeed a future outer trapping horizon. Therefore, the respective observers see it as a cosmological black hole, contrary to some claims in the literature.

\end{abstract}

\maketitle


\section{Introduction}
\label{sec1}
In recent papers \cite{Boehm:2020jwd, Picker:2021jxl,  Boehm:2021kzq}, we have advocated that a realistic description of primordial black holes (PBHs) must incorporate the effects of cosmological expansion prior to their decoupling from the Hubble flow. Notably, the Schwarzschild black hole, which is asymptotically flat and not surrounded by the cosmic fluid, cannot be an adequate description of black holes in the early universe. Among several alternatives for cosmological black holes \cite{McVittie:1933zz, Einstein:1945id, Lemaitre:1933gd, Tolman:1934za, Bondi:1947fta, thakurta} discussed in the literature (see Ref. \cite{Faraoni:2015ula} for a review), we have argued that the Thakurta black hole \cite{thakurta} is a particularly viable candidate. Most importantly, the Thakurta metric demonstrates the significant phenomenological effects which may arise from properly accounting for cosmological embeddings of black  holes. Adopting the Thakurta metric, we have found that the formation rate of primordial black hole binaries are slowed down by the cosmic expansion \cite{Boehm:2020jwd} and that cosmological black holes evaporate faster than their Schwarzschild counterparts \cite{Picker:2021jxl}. These effects significantly alter the established bounds on primordial black hole abundance \cite{Carr:2020gox}. Indeed, since our initial work, there seems to be increased interest in the phenomenological effects of black hole effective masses due to cosmological contributions~\cite{Xavier:2021chn,Croker:2021duf}.

There has been recent discussion in the literature over the validity of the Thakurta metric as a black hole solution~\cite{Hutsi:2021nvs,Boehm:2021kzq}. One issue regards the nature of the Thakurta apparent horizon. In Ref.~\cite{Harada:2021xze}, the authors claim that the horizon of the Thakurta spacetime is a past trapping horizon, and therefore the Thakurta horizon is either that of a white hole or a cosmological horizon, but not a black hole (see also the earlier work~\cite{Mello:2016irl}). 

In this work, we reexamine the causal structure of the Thakurta spacetime. We note that the expansion scalars which define the nature of the horizon, although coordinate-independent, do still depend explicitly on the choice of spacetime foliation~\cite{Abreu:2010ru, Faraoni:2016xgy}. Recall that stationary black holes admit a time-like Killing vector and so the preferred foliation (and thus time evolution) is uniquely given by the Cauchy surfaces. However, for dynamical black holes, no such time-like Killing vector exist, and hence the time evolution is not uniquely defined. Namely, for an apparent horizon, at any given time its evolution depends on the chosen direction of time flow at each given spatial point. The stack of such spatial surfaces with prescribed time flow (i.e, the foliation) identifies a class of observers that would describe a specific time evolution and the nature of the apparent horizon. This description can differ for different foliations. 

For spherically symmetric non-stationary spacetimes, there is a geometrically-preferred foliation known as the Kodama foliation~\cite{Kodama:1979vn}. The time flow for this foliation is defined by the Kodama vector, which is covariantly conserved, similar to the Killing vector for stationary spacetimes. In what follows, we explicitly demonstrate that observers associated with the Kodama foliation observe the Thakurta horizon as the future outer trapping horizon and thus would describe the spacetime as a cosmological black hole. With this inescapable ambiguity in mind, the Thakurta spacetime is a perfectly legitimate solution for a black hole embedded in expanding universe.   



\section{The Thakurta horizons}
\label{sec2} 
The non-rotating Thakurta line element can be written in `cosmological' coordinates as,
\begin{eqnarray}
ds^2&=&f(r)dt^2-\frac{a^2(t)}{f(r)}dr^2 -a^2(t)r^2d\Omega_2~,
\label{eq1}
\end{eqnarray}
where $d\Omega_2=d\theta^2+\sin^2\theta d\phi^2$, $a$ is the scale factor of the expanding universe, $t$ is the cosmic times and $f(r)=1-2Gm/r\equiv 1-2Gma/R$ and $R=a r$ is the areal radial coordinate.  

The spacetime (\ref{eq1}) interpolates between the spatially flat Friedmann–Lemaître–Robertson–Walker spacetime as $f(r\to \infty)\to 1$ and the Schwarzschild spacetime as the Hubble expansion rate $H\equiv \frac{d\ln a}{dt}\to 0$. The Thakurta metric exhibits two apparent horizons, which can be interpreted respectively as the cosmological and black hole horizons,
\begin{eqnarray}
\label{h1}
R_{ch}=\frac{1}{2H}\left(1+\sqrt{1-8HGma}\right) \\
R_{bh}=\frac{1}{2H}\left(1-\sqrt{1-8HGma}\right)
\label{h2}
\label{hor}
\end{eqnarray}
We note that in the static limit $H\to 0$,  the cosmological horizon $R_{ch}\to \infty$ and the black hole apparent horizon approaches the Schwarzschild event horizon $R_{bh}=2Gm$ ($a=1$).  In the limit of vanishing inhomogeneity, $m\to 0$ or in the very early universe $a\to 0$, the black hole horizon  $R_{bh}\to 0$, and the cosmological horizon approaches the Hubble horizon, $R_{ch}\to 1/H$.  

The nature of apparent horizons is known to be observer-dependent \cite{Figueras:2009iu, Faraoni:2016xgy}. Indeed, on top of the local geometry, one needs to specify also a foliation of the spacetime, that is, to define a series of spatial slices that are simultaneous for a class of observers. There is no unique way to define the foliation, since the direction of time at each given spatial point is ambiguous. This is of course related to the fact that, unlike for stationary black holes, we no longer have a time-like Killing vector that would define the time evolution of dynamical black holes unambiguously. Fortunately, for spherically symmetric dynamical black holes, there is a  preferred foliation \cite{Abreu:2010ru} that is associated with the Kodama time \cite{Kodama:1979vn}. The Thakurta metric in Kodama time $\tau$ and Schwarzschild-like coordinates takes the form,
\begin{align}
&ds^2=F(R,t)~d\tau^2-
\frac{dR^2}{F(R,t)} - R^2d\Omega_2~,\nonumber\\
&F(R,t) = \left(1-\frac{2Gm_{MS}}{R}\right)~,
\label{eq2}
\end{align}
where $m_{MS}$ is the quasi-local Misner-Sharp mass \cite{Misner:1964je} which for the Thakurta metric reads,
\begin{equation}
m_{MS}=ma+\frac{H^2R^3}{2Gf(R)}~.
\label{ms}
\end{equation}
Two horizons in Eqs. (\ref{h1}) and (\ref{h2}) both satisfy the following equation:
\begin{equation}
R_h=2Gm_{MS}(R_h)~,
\label{h}
\end{equation}

Following \cite{Picker:2021jxl}, from Eq.~(\ref{eq2}) we consider here the gauge where the Kodama time translation vector coincides with the Kodama vectors. The Kodama time $\tau$ is then related to the cosmic time $t$ through the following equation \cite{Picker:2021jxl}:
\begin{equation}
d\tau = dt+\frac{HR}{f(R)}\frac{dR}{F(R,t)}~.
\label{time}
\end{equation} 


Apparent horizons are defined locally using congruences of null geodesics. Hence, to define their nature we must compute the expansion scalars $\theta_{\ell}$ and $\theta_{n}$ for outgoing and incoming null geodesics, respectively~\cite{Hayward:1993wb}. The general expressions for these expansion scalars are,
\begin{eqnarray}
\label{eql}
\theta_\ell& =& \nabla_{\mu}\ell^{\mu}+\ell^{\mu}\ell_{\nu}\nabla_{\mu}n^{\nu}~, \\
\theta_n & =& \nabla_{\mu}n^{\mu}+n^{\mu}n_{\nu}\nabla_{\mu}\ell^{\nu}~, 
\label{eqn}
\end{eqnarray}
where $\ell^{\mu}$ and $n^{\mu}$ are tangent vectors to outgoing and incoming null radial geodesics, $\ell^{\mu}\ell_{\mu}=n^{\mu}n_{\mu}=0$. Here we use the cross-normalisation $\ell^{\mu}n_{\mu}=1$. An apparent horizon is considered a black hole horizon if it is future (marginally) trapped, i.e., $\theta_{\ell}=0$ (at the horizon) and $\theta_{n}<0$. Physically the first conditions implies that  the future-pointing outgoing null rays momentarily stop propagating outward and, presumably, turn around at the horizon---the outward propagating light is dragged back by the strong gravity. The second condition implies that future-directed incoming null rays always converge inward, unlike white holes for which $\theta_{n}\geq 0$. For cosmological trapped horizons where the observer is surrounded by the horizon we have instead, $\theta_n=0$ (at the horizon) and $\theta_{\ell}>0$. We will show explicitly that for cosmological foliation (\ref{eq1}) the Thakurta horizons are seen as either cosmological or white hole horizons in accord with the results obtain in Refs.~\cite{Harada:2021xze,Mello:2016irl}. However, the Kodama observers (\ref{eq2}) see Thakurta horizons as describing black holes.

\textbf{The cosmological foliation.} For the cosmological foliation (\ref{eq1}) the null radial geodesics ($ds^2=d\Omega_2=0$) are given by,
\begin{equation}
  \frac{dr}{dt}=\pm \frac{f}{a}~.
\end{equation}
Consequently the tangent vectors read:
\begin{eqnarray}
\label{l1}
\ell^{\mu}& =& \left(1, f/a,0,0\right)~, \\
n^{\mu} & =& \frac{1}{2f}\left(1, -f/a,0,0\right)~, 
\label{n1}
\end{eqnarray}

Using Eqs. (\ref{eql},\ref{eqn}) and the Christoffel symbols from Appendix, we obtain,
\begin{eqnarray}
    \theta_{\ell}&=&\frac{2}{R}\left(HR+f(R)\right)~, \\
    \theta_{n}&=&\frac{1}{Rf(R)}\left(HR-f(R)\right)~. 
\end{eqnarray}
At the Thakurta horizons $HR=f(R)$, $\theta_{n} =0$ and $\theta_{\ell}>0$. Thus, according to the 
cosmological observers, the Thakurta horizons are 
seen as past apparent horizons, because the light rays exiting the horizons never return back. To determine whether 
they are also trapping horizons, we compute,   
\begin{eqnarray}
    \left. \mathcal{L}_{\ell}\theta_n\right\vert_{R=R_{h}}&=&\left.\left(\partial_t+\frac{f}{a}\partial_r\right)\theta_n\right\vert_{R=R_{h}} \nonumber \\
    &=&\left.-\frac{1}{HR}\left(\frac{4\pi G}{3}f(\rho-3p)+\frac{H\left(1+HR\right)}{R}\right)\right\vert_{R=R_{h}}~,
    \end{eqnarray}
where we have used the trace of the Einstein equations, $\dot H+2H^2=-\frac{8\pi G}{3}f(\rho-3p)$, to rewrite the last expression in terms of density ($\rho$) and pressure ($p$) of the cosmological fluid\footnote{The Thakurta metric is supported by an imperfect fluid with non-zero radial heat flow. In addition, there is assumed to be no radial fluid accretion---that is, the fluid 4-velocity is orthogonal to the 4-vector that defines the heat flow. The cosmological observers are then those who observe the cosmic fluid at rest.} evaluated at $R=R_h$. It is obvious that the above expression is negative definite as soon as $\rho\geq 3p$, just as in a radiation or matter dominated universe. Hence, the Thakurta horizons are seen by the cosmological observers as past outer trapped horizons. That is, they would described the trapped regions $R< R_{bh}$ and $R>R_{ch}$ as the white hole and cosmologically trapped regions. Note that a region $R_{bh}<R<R_{ch}$ is not trapped, since $\theta_{\ell}\theta_{n}<0$ there.

\textbf{The Kodama foliation.} We will show now that Kodama observers see the Thakurta horizons differently. The Schwarzschild-like coordinates (\ref{eq1}) for the Kodama foliation exhibit coordinate singularities at the Thakurta horizons (\ref{h}). Therefore, we will work in Painlev\'e-Gullstrand coordinates instead, which are well suited for future (but not for past) horizons. The line element (\ref{eq1}) in those coordinates is given by,
\begin{equation}
    ds^2=F~d \tilde\tau^2-\sqrt{\frac{8Gm_{MS}}{R}}d\tilde\tau dR -
dR^2 - R^2d\Omega_2~,
    \label{eq3}
\end{equation}
where the new time coordinate $\tilde \tau$ is defined as: $\mathrm{d} \tilde \tau = \mathrm{d}\tau+ \mathrm{d}R\sqrt{1-F}/F$. From the above equation we find that the radial null rays satisfy,
\begin{equation}
    \frac{dR}{d\tilde\tau}= -\sqrt{\frac{2Gm_{MS}}{R}}\pm 1.
    \label{rays1}
\end{equation}
The outgoing and incoming tangent vectors are then,
\begin{eqnarray}
\label{l2}
\ell^{\mu}& =& \left(1, 1-\sqrt{2Gm_{MS}/R},0,0\right)~, \\
n^{\mu} & =& \frac{1}{2}\left(1, -1-\sqrt{2Gm_{MS}/R},0,0\right)~, 
\label{n2}
\end{eqnarray}
The calculation of the expansion scalars using the Christoffel symbols displayed in the Appendix gives us,
\begin{eqnarray}
    \theta_{\ell}&=&\frac{2}{R}\left(1-\sqrt{1-F}\right)~, \\
    \theta_{n}&=&-\frac{1}{R}\left(1+\sqrt{1-F}\right)~. 
\end{eqnarray}
We observe that at the Thakurta horizons (see Eq. (\ref{h})) $F=0$ and hence, $\theta_{\ell}=0$ and $\theta_n<0$. This implies that, unlike the cosmological observers, the Kodama observers see the Thakurta horizons as future apparent horizons. Further we compute the following:
\begin{align}
    \left. \mathcal{L}_{n}\theta_\ell\right\vert_{R=R_{h}} &= 
     \left.\frac{1}{2R}\left(\dot F-2F^{\prime}\right)\right\vert_{R=R_{h}}\nonumber\\
     &= \left. -4\pi G \frac{\dot R_h+2}{\dot R_h}~T_{\mu\nu}\ell^{\mu}\ell^{\nu}\right\vert_{R=R_{h}}
\label{grad}
\end{align}
where we have used $F(R_h(\tilde \tau),\tilde \tau)=0$ and the Einstein equations, $-\frac{\dot F}{R_h}=G_{\mu\nu}\ell^{\mu}\ell^{\nu}=8\pi G T_{\mu\nu}\ell^{\mu}\ell^{\nu}$ evaluated at $R=R_h$,  
to arrive at the second line. Next, if we take the full $\tilde \tau$ derivative of Eq. (\ref{h}) and use $\dot m_{MS}=\left. 4\pi R_h^2 T_{\mu\nu}\ell^{\mu}\ell^{\nu}\right\vert_{R=R_h}$, we arrive at
\begin{eqnarray}
\dot R_h&=&\left.\frac{2G\dot m_{MS}}{1-2Gm^{\prime}_{MS}}\right\vert_{R=R_h} \nonumber \\
&=&\left. 2G\frac{4\pi R_h^2 T_{\mu\nu}\ell^{\mu}\ell^{\nu}}{1-2Gm^{\prime}_{MS}}\right\vert_{R=R_h}~.
\end{eqnarray}
For a region of interest $R\geq R_h=2Gm_{MS}$, we have $1-2Gm^{\prime}_{MS}\geq 0$. Therefore, under the null energy condition $T_{\mu\nu}\ell^{\mu}\ell^{\nu}\geq 0$, the evolving Thakurta horizon is a non-decreasing function of the Kodama time, $\dot R_h\geq 0$. This in turn implies that $\mathcal{L}_{n}\theta_\ell\vert_{R=R_{h}}<0$ (see Eq. (\ref{grad})). Thus, the Thakurta horizon is seen as a future outer trapped horizon by the Kodama observers, consistent with black hole interpretation of the metric. Our analysis is consistent with the one in Ref. \cite{Nielsen:2005af}.   

\section{Conclusion}
\label{sec5}
The characteristics of time-dependent black hole horizons depends on the spacetime foliation. That is, observers associated with different foliations may perceive different horizon evolution. In this paper, we have shown explicitly that the horizon of the Thakurta spacetime differs between the cosmological foliation, and the Kodama foliation. The cosmological observers perceive a white hole/cosmological horizon, while observers in the Kodama basis perceive a black hole. 

Such an ambiguity in time evolution is common for time-dependent spacetimes and should not be a cause of confusion. The key reason for this is the absence of a time-like Killing vector, and thus an inability to assign a universal time flow. Nevertheless, for spherically symmetric spacetimes, there is a preferred Kodama foliation along with an associated Kodama time, which appears to be the most suitable choice for the description of evolving horizons. In particular, we have explicitly demonstrated that the Thakurta horizon in this foliation is indeed a future outer trapping horizon. We conclude, therefore, that the Thakurta metric is a legitimate solution for a cosmological black hole, and further studies of their distinct phenomenology remain important as we continue to understand the role of primordial black holes in the early universe.



\begin{acknowledgments}
The work of AK was partially supported by the Australian Research Council through the Discovery Project grant DP210101636 and by the Shota Rustaveli National Science Foundation of Georgia (SRNSFG) through
the grant DI-18-335
\end{acknowledgments}

\appendix*
\section{Appendix:~ The Christoffel symbols}
For convenience, we provide explicit Christoffel symbols used in our calculations. 

For the metric (\ref{eq1}) the non-vanishing Christoffel symbols are:
\begin{eqnarray*}
\Gamma^{0}_{01}&=&\Gamma^{0}_{01}\frac{a^2}{f^2}=-\Gamma^{1}_{11}=\frac{f^{\prime}}{2f}~,\\
\Gamma^{0}_{11}&=&\Gamma^{0}_{22}\frac{1}{r^2f}=\Gamma^{0}_{33}\frac{1}{r^2\sin^2\theta f}=\frac{a^2}{f^2}H~, \\
\Gamma^{1}_{01}&=&\Gamma^{2}_{02}=\Gamma^{3}_{03}=H,~\Gamma^{1}_{22}=\Gamma^{1}_{33}\frac{1}{\sin^2\theta}=-rf \\
\Gamma^{2}_{21}&=&\Gamma^{3}_{31}=\frac{1}{r},~\Gamma^{2}_{23}=-\Gamma^{3}_{32}\sin^2\theta =-\sin\theta\cos\theta~.
\end{eqnarray*}

For the metric (\ref{eq3}) the non-vanishing Christoffel symbols are:
\begin{eqnarray*}
\Gamma^{0}_{00}&=&\Gamma^{0}_{01}\sqrt{1-F}
=\Gamma^{0}_{11}\left(1-F\right)= -\Gamma^{1}_{01} 
=- \Gamma^{1}_{11}\sqrt{1-F}\\
&=&\frac{G}{R^2}\sqrt{1-F}
\left(m_{MS}-Rm^{\prime}_{MS}\right)~, \\
\Gamma^{1}_{00}&=&\frac{Gm_{MS}}{R^2}\left(1-\frac{G}{R}\right)\left(m_{MS}-m^{\prime}_{MS}\right) \\
&-&\frac{Gm_{MS}}{R}\left(\frac{Gm^2_{MS}}{R^2}-\frac{\dot m_{MS}}{m_{MS}}\sqrt{\frac{R}{2Gm_{MS}}}\right)~, \\
\Gamma^{0}_{22}&=&\Gamma^{0}_{33}\frac{1}{\sin^2\theta}=-\sqrt{1-F}R~, \\
\Gamma^{1}_{22}&=&\Gamma^{1}_{33}\frac{1}{\sin^2\theta}=-F R~,\\
\Gamma^{2}_{21}&=&\Gamma^{3}_{31}=\frac{1}{R},~\Gamma^{2}_{23}=-\Gamma^{3}_{32}\sin^2\theta =-\sin\theta\cos\theta~.
\end{eqnarray*}
In the above equations, $X^\prime \equiv \frac{\partial X}{\partial r}$ or $\frac{\partial X}{\partial R}$ 
and $\dot X \equiv \frac{\partial X}{\partial t}$ or $\frac{\partial X}{\partial \tilde \tau}$.

\bibliography{bib.bib}
\bibliographystyle{bibi}

\end{document}